\documentclass[twocolumn,aps,showpacs,floatfix,superscriptaddress]{revtex4}
\usepackage{amsmath,amssymb,eucal,graphicx}
\begin{document}
\title{Addition-Deletion Networks}
\author{E.~Ben-Naim}
\email{ebn@lanl.gov}
\affiliation{Theoretical Division and Center for Nonlinear Studies,
Los Alamos National Laboratory, Los Alamos, New Mexico 87545, USA}
\author{P.~L.~Krapivsky}
\email{paulk@bu.edu}
\affiliation{Department of Physics, Boston University, Boston, 
Massachusetts 02215, USA}
\begin{abstract}
  We study structural properties of growing networks where both
  addition and deletion of nodes are possible.  Our model network
  evolves via two independent processes. With rate $r$, a node is
  added to the system and this node links to a randomly selected
  existing node. With rate $1$, a randomly selected node is deleted,
  and its parent node inherits the links of its immediate descendants.
  We show that the in-component size distribution decays
  algebraically, $c_k\sim k^{-\beta}$, as $k\to\infty$. The exponent
  $\beta=2+(r-1)^{-1}$ varies continuously with the addition rate
  $r$. Structural properties of the network including the height
  distribution, the diameter of the network, the average distance
  between two nodes, and the fraction of dangling nodes are also
  obtained analytically. Interestingly, the deletion process leads to
  a giant hub, a single node with a macroscopic degree whereas all
  other nodes have a microscopic degree.

\end{abstract}
\pacs{89.75,Hc, 05.40.-a,05.20.Dd, 02.50.Ey}
\maketitle

\section{Introduction}

Idealized models of random networks provide an important tool for
characterizing real-world networks such as communication,
infrastructure, and social networks \cite{ws,ab,dm}. For example, the
emergence of hubs in heterogeneous networks is a consequence of a
``rich-gets-richer'' mechanism. This phenomenon arises in networks
that grow via preferential attachment \cite{ba,krl,dms}.

Most network growth models are based upon sequential addition of nodes
and subsequent attachment to existing nodes according to a prescribed
mechanism. This framework does not allow for removal of nodes.  In
reality, there are two types of networks. One type of networks
including for example citation networks may only expand, while another
type of networks such as friendship networks may either expand or
contract. Indeed, individuals may join or leave a social group
\cite{deb,nowak}.  The same is true for technological networks such as
the world-wide-web because websites may disappear.

There is substantial theoretical understanding of the structure of
strictly expanding networks, but much less is known about networks
that may also contract \cite{deb,nowak,mgp,cl,cfv,mgn,fc}. In the
latter case, it is typically impossible to construct closed equations
for quantities such as the degree distribution because removal of
nodes generates memory or correlations.

In this paper, we introduce a simple network growth process with
addition and deletion of nodes and show that it is possible to obtain
closed equations for several structural properties of the network.  In
our growing network model, nodes may be added or deleted. When a node
is added to the network, it is attached to a randomly selected
existing node. When a node is deleted, its daughters are attached to
its parent. This process is relevant for phylogenetic trees, namely
trees that document the ancestries of distinct biological species
\cite{mk,msw,dekm} because in the process of evolution, mutation
events generate new nodes while extinction events result in deletion
of existing nodes \cite{hv,kwk,mz,mal,bk}.

We first investigate the in-component size distribution and show that
this quantity obeys a closed evolution equation. We analytically
obtain many characteristics of this distribution including the
generating function, the moments, and the tail behavior. The
in-component size distribution decays algebraically and the
characteristic exponent varies continuously with the addition
rate. Such behavior is in contrast with preferential attachment
networks where this exponent is fixed.

Next, we obtain closed equations for the height distribution that
characterizes the distance between a node and the root. This
distribution obeys Poisson statistics. Other structural
characteristics such as the diameter of the network and the degree of
the most connected node follow directly from this quantity. In
general, deletion leads to a condensation phenomena where a single
node is connected to a finite fraction of all nodes, while the rest of
the nodes have only a finite degree.

The rest of this paper is organized as follows. The addition-deletion
process is introduced in section II. The in-component size
distribution and related quantities such as the fraction of dangling
nodes are discussed in section III. Next, in section IV, the height
distribution and related quantities including the diameter are
described.  The degree distribution is analyzed using an approximate
theory and numerical simulations in section V.  Section VI contains a
brief summary. Finally, appendices \ref{Alpha} and \ref{Aggr} contain
several technical derivations.

\section{The Addition-Deletion Process}

In the addition-deletion process, initially there is one seed node,
and then, the network grows via addition of nodes and shrinks via
deletion of nodes as follows.
\begin{enumerate}
\item {\bf The addition process.} With rate $r$, a new node is added
and it links to a randomly selected node (Fig.~1). This is an
egalitarian attachment process as the new node is equally likely to
link to any one of the existing nodes.
\item {\bf The deletion process.} With rate $1$, a randomly selected
node (together with its outgoing link) is deleted. The parent of the
deleted node inherits all the incoming links of the deleted node
(Fig.~2). This inheritance mechanism preserves ancestral relations as
is the case in phylogenetic trees: when a species goes extinct, its
immediate descendants are linked to its immediate ancestor.
\end{enumerate}
We stress that the addition process and the deletion process are
completely independent of each other and that in the deletion process
all links are removed with an equal probability.

\begin{figure}[t]
\includegraphics[width=0.4\textwidth]{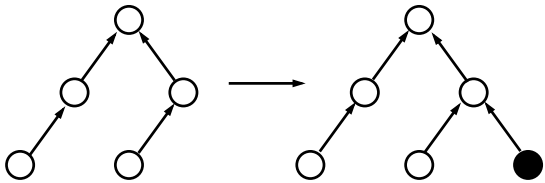}
\caption{Illustration of the node addition process.}
\label{fig-addition}
\end{figure}

This addition-deletion process preserves the complexity of the graph
since nodes and links are added and deleted in pairs. Thus, starting
with a single seed node and no links, the network has a tree
structure. The seed node can not be deleted (or the ancestry would be
destroyed) and therefore, the seed is the root of the tree.

\begin{figure}[t]
\includegraphics[width=0.4\textwidth]{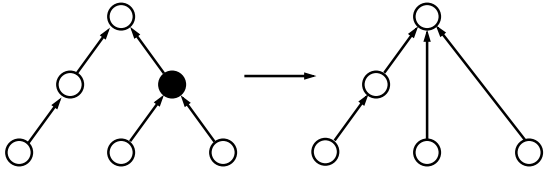}
\caption{Illustration of the node deletion process.}
\label{fig-deletion}
\end{figure}

The average total number of nodes at time $t$, $N(t)$, evolves
according to $dN/dt=(r-1)$. Therefore, this quantity grows linearly
with time and since $N(0)=1$,
\begin{equation}
\label{nt}
N(t)=1+(r-1)t.
\end{equation}
The total number of links, $L(t)$, follows from $L=N-1$ and therefore,
$L(t)=(r-1)t$. We restrict our attention to the case of growing
networks, $r>1$. In this case, the size of the network increases
indefinitely with time and fluctuations are negligible in the long
time limit. The total number of nodes grows diffusively and is subject
to large fluctuations in the limit $r\to 1$.

\section{The in-component size distribution}

The in-component of a node ${\bf n}$ is the set of all nodes which are
connected to ${\bf n}$ via a path of directed links. In the context of
a phylogenetic tree, the in-component is simply the set of all
descendants.  Let $C_k$ be the average number of nodes with an
in-component of size $k$.  In our definition, the in-component
includes the node itself, and therefore $k\geq 1$.  The in-component
size distribution evolves according to
\begin{eqnarray}
\label{Ck-eq}
\frac{dC_k}{dt}\!=\!r\delta_{k,1}\!+\!
\frac{r}{N}\left[(k\!-\!1)C_{k-1}\!-\!kC_k\right]
\!+\!\frac{k}{L}\left[C_{k+1}\!-\!C_k\right].
\end{eqnarray}
The first three terms correspond to changes caused by node
addition. New nodes generate a new in-component of size one and thus,
the first term.  In-components grow by one when a node links to any of
the nodes in that in-component and the next two terms account for this
augmentation. The corresponding attachment rate is the addition rate
$r$ normalized by the total number of links $N$.  The last two terms
correspond to node deletion. Deletion of any one of the $k$ links in
an in-component of size $k+1$ generates an in-component of size $k$
and hence the gain term. Similarly, deletion of any one of the $k$
nodes in an in-component of size $k$ decreases the in-component size
by one and hence the loss term. All nodes except for the root may be
deleted and accordingly, the deletion rate is normalized by the total
number of links $L$. By summing equations (\ref{Ck-eq}) one can verify
that the total number of nodes $N=1+\sum_{k\geq 1}C_k$ satisfies
$dN/dt=r-1$.

We are interested in the long time asymptotics where the difference
between the total number of nodes and the total number of links is
negligible.  Therefore, we replace $L$ with $N$ in (\ref{Ck-eq}). We
also make the transformation
\begin{equation}
\label{ck-def}
C_k=Nc_k
\end{equation}
where $c_k$ is the fraction of nodes with in-component degree equal to
$k$ and $\sum_{k\geq 1} c_k=1$.  The distribution function $c_k$ satisfies the
difference equation
\begin{eqnarray}
\label{ck-eq}
(r\!-\!1)c_k\!=\!r\left[\delta_{k,1}\!+\!(k\!-\!1)c_{k-1}\!
-\!kc_k\right]\!+\!k(c_{k+1}\!-\!c_k).
\end{eqnarray}
The tail of this distribution can be obtained using a continuum
approximation, that is, through replacement of differences with
derivatives: $c_{k+1}-c_k\to dc/dk$, etc. This change transforms the
difference equation (\ref{ck-eq}) into the differential equation
\begin{equation}
\label{ck-eq-cont}
r\frac{d}{dk}(k\,c)-k\frac{dc}{dk}+(r-1)c=0.
\end{equation}
The solution of this equi-dimensional equation is a power-law
\begin{equation}
\label{beta}
c_k\simeq B\,k^{-\beta}\,,\qquad \beta=2+\frac{1}{r-1},
\end{equation}
as $k\to\infty$.  Thus, the in-component size distribution has a
power-law tail. The exponent $2\leq \beta < \infty$ varies
continuously with the addition rate. The characteristic exponent is
minimal, $\beta=2$, for the well-understood case of random recursive
trees ($r\to\infty$) where deletion is irrelevant
\cite{hmm,dek,mh,drmota,kr1,kr2}.  The exponent diverges in the more
interesting limit of marginally growing networks ($r\to 1$) where
deletion is as strong as addition.

Addition-deletion networks differ from strictly growing networks where
the exponent characterizing the in-component size distribution is
independent of the details of the attachment mechanism, $\beta=2$
\cite{kr1}. Thus the deletion process qualitatively changes the
network structure.

The moments \hbox{$M_n=\sum_k k^n C_k$} provide complementary
information about the in-component size distribution. Of course,
$M_0=N-1$. The first moment satisfies
\begin{equation}
\label{M1-eq}
\frac{dM_1}{dN}=\frac{r-2}{r-1}\,\frac{M_1}{N}+\frac{r+1}{r-1},
\end{equation}
as seen by summing \eqref{Ck-eq}.  Equation \eqref{M1-eq} describes
the evolution as a function of the total number of nodes rather than
time. Since we are interested in the asymptotic behavior, we again
replace $L$ with $N$ in (\ref{Ck-eq}).  Solving Eq.~(\ref{M1-eq}), we
find that the first moment is proportional to the number of nodes,
\hbox{$M_1=(1+r)N$}. Hence the average in-component size is linear in
the addition rate, \hbox{$\langle k\rangle=1+r$}.

Similarly, the second moment obeys 
\begin{equation}
\frac{dM_2}{dN}=\frac{2r-3}{r-1}\,\frac{M_2}{N}+
\frac{r+3}{r-1}\,\frac{M_1}{N}+1.
\end{equation}
There are three types of asymptotic behavior
\begin{equation}
\label{M2}
M_2\simeq 
\begin{cases}
\frac{(r^2+5r+2)}{2-r}N            &1<r<2;\\
16\, N\ln N                               &r=2;\\
\kappa N^{\frac{r-2}{r-1}}         &r>2.
\end{cases}
\end{equation}
The proportionality constant $\kappa$ depends on the initial
conditions. These asymptotic behaviors are consistent with the
algebraic decay (\ref{beta}): the second moment grows linearly when
$\beta>3$, it increases super-linearly when $\beta<3$, and it acquires
a logarithmic correction in the marginal case $\beta=3$.

The generating function 
\begin{equation}
F(x)=\sum_{k=1}^\infty c_k x^k 
\end{equation}
contains complete information about the in-component size
distribution.  Normalization implies $F(1)=1$ and since $c_0=0$, then
$F(0)=0$. Multiplying Eq.~(\ref{ck-eq}) by $x^k$ and summing over $k$, 
we arrive at a first order differential equation for the generating
function
\begin{equation}
\label{Fx-eq}
(x-1)(rx-1)F'(x)+\left(1-r-x^{-1}\right)F(x)+rx=0
\end{equation}
where the prime denotes differentiation with respect to $x$.  The
homogeneous part of equation \eqref{Fx-eq} admits the solution 
$x(1-x)^{\beta -1}|rx-1|^{-\beta}$.  We make the transformation
\hbox{$F(x)=x(1-x)^{\beta-1}(rx-1)^{-\beta}\,a(x)$} when
\hbox{$1/r<x<1$}, and then the auxiliary function $a(x)$ satisfies
\hbox{$a'=r\,(1-x)^{-\beta}(rx-1)^{\beta-1}$}. Integration of this
equation yields the generating function
\begin{equation}
\label{Fx-sol-a}
F(x)=
r\,\frac{x(1-x)^{\beta-1}}{(rx-1)^\beta}
\int_{1/r}^x dy\,\frac{(ry-1)^{\beta-1}}{(1-y)^\beta}
\end{equation}
in the range $1/r<x<1$.  The lower limit of integration was chosen to
assure that $F(1/r)$ is finite. One can verify that \hbox{$\lim_{x\to
1}F(x)=1$} in agreement with the normalization requirement.  A similar
calculation gives the generating function in the range $0<x<1/r$
\begin{equation}
\label{Fx-sol-b}
F(x)=r\,\frac{x(1-x)^{\beta-1}}{(1-rx)^\beta}
\int_x^{1/r} dy\,\frac{(1-ry)^{\beta-1}}{(1-y)^\beta}.
\end{equation}

The large-$k$ behavior is encoded in the $x\to 1$ behavior of the
generating function. First, we confirm the power-law tail (\ref{beta})
by using (\ref{Fx-sol-a}). Second, we obtain the corresponding
proportionality constant
\begin{equation}
\label{alpha}
B=\Gamma(\beta)(\beta-1)^\beta.
\end{equation}
The appendix details the analysis leading to this result.

\begin{figure}[t]
\includegraphics[width=0.45\textwidth]{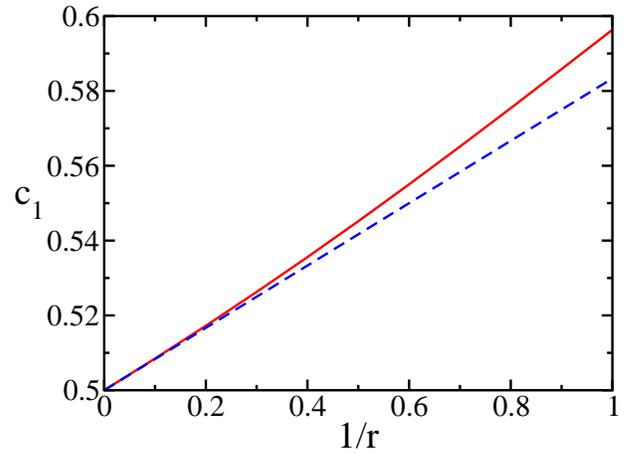}
\caption{The fraction of dangling nodes $c_1$ versus inverse of the
addition rate $1/r$ (solid line). Also shown is the large-$r$
asymptotic behavior $c_1=\frac{1}{2}+\frac{1}{12r}$ (dashed line).}
\label{fig-dangling}
\end{figure}

Conversely, the small-$k$ behavior is reflected by the $x\to 0$
behavior of the generating function. The fraction of dangling nodes,
that is, nodes with no incoming links, is given by $c_1=\lim_{x\to
0}F'(x)$ and by differentiation of the generating function
(\ref{Fx-sol-b}) we readily find
\begin{equation}
\label{c1-sol}
c_1=\int_0^{1} dy\,(1-y)^{\beta-1}(1-y/r)^{-\beta}.
\end{equation}
The integral can be expressed via the hypergeometric function,
$c_1=\beta^{-1}F(\beta,1;\beta+1;r^{-1})$, or alternatively as the
power series $c_1=\sum_{k\geq 0}(k+\beta)^{-1}\,r^{-k}$. The quantity
$c_1$ is a monotonically decreasing function of the addition rate $r$
and is roughly linear in $1/r$ (figure \ref{fig-dangling}).  It is
maximal for marginally growing networks (see subsection A below),
\begin{equation}
\label{c1-sol-1}
\lim_{r\to 1}c_1=e\,E_1(1)=0.596347\ldots,
\end{equation}
where $E_1(x)=\int_x^\infty dz z^{-1}e^{-z}$ is the exponential
integral \cite{as}.  The fraction of dangling nodes is minimal for
random recursive trees, $\lim_{r\to \infty}c_1=1/2$. In general, over
half of the nodes are dangling nodes.

The in-component size distribution can be calculated recursively for
small $k$ by using \eqref{ck-eq}, \hbox{$c_2=r(2c_1-1),
c_3=r\left[(6r+1)c_1-(3r+1)\right]$}, etc.

We now investigate the limits $r\to 1$ and $r\to \infty$ in detail.  

\subsection{Marginally Growing Networks}

The recursion relation (\ref{ck-eq}) becomes
\begin{equation}
\label{ck-eq-1}
c_{k+1}=\delta_{k,1}+(k-1)c_{k-1}-2kc_k+(k+1)c_{k+1}
\end{equation}
for marginally growing networks ($r\to 1$).  Repeating the steps
leading to (\ref{ck-eq-cont}) yields the second order linear
differential equation 
\begin{equation}
\label{ck-eq-cont-1}
\frac{d^2}{dk^2}\left(kc\right)-\frac{dc}{dk}-c=0.
\end{equation}
The tail of the in-component size distribution has a stretched
exponential tail
\begin{equation}
\label{ak-asympt}
c_k\sim k^{-1/4}\,e^{-2\sqrt{k}}, 
\end{equation}
a result that can be obtained using the WKB technique.  This sharp
decay is consistent with the divergent $\beta$.

To find the generating function, we repeat the steps leading to
(\ref{Fx-sol-a}).  The governing equation (\ref{Fx-eq}) becomes
$(x-1)^2F'-x^{-1}F+x=0$, and the solution of this equation is
\hbox{$F(x)=\frac{x}{1-x}\,e^{\frac{1}{1-x}} \int_0^{1-x}
\frac{dy}{y}\,\exp\big(-\frac{1}{y}\big)$}.  The generating function
can be expressed in terms of the exponential integral
\begin{equation}
F(x)=\frac{x}{1-x}\,e^{\frac{1}{1-x}}E_1\left(\frac{1}{1-x}\right),
\end{equation}
{}from which the fraction of dangling nodes (\ref{c1-sol-1}) follows.

\subsection{Random Recursive Trees}

The second limiting case, $r\to\infty$, corresponds to random
recursive trees. Now, the in-component size distribution satisfies the
recursion equation
\begin{equation}
(k+1)c_k=(k-1)c_{k-1}+\delta_{k,1}.
\end{equation}
Unlike in the general case, this simpler equation can be solved
recursively starting at $k=1$: $c_1=\frac{1}{1\cdot 2}$,
$c_2=\frac{1}{2\cdot 3}$, $c_3=\frac{1}{3\cdot 4}$, etc. We thus
recover the well-known result \cite{kr1}
\begin{equation}
\label{ck-sol-2}
c_k=\frac{1}{k(k+1)}.
\end{equation}
Alternatively, one can obtain the distribution from the generating
function $F(x)=1+\frac{1-x}{x}\,\ln(1-x)$.

\section{The Height Distribution}

Another important structural characteristic is the height of a node,
that is, the number of links separating a given node from the root
\cite{bkm,ra}.  Let $H_k(t)$ be the average number of nodes with
height $k$ at time $t$. This quantity evolves according to
\begin{equation}
\label{Hkt-eq}
\frac{dH_k}{dt}=\frac{r}{N}H_{k-1}+\frac{k}{L}\,\left(H_{k+1}-H_k\right)
\end{equation}
for $k\geq 1$. The initial condition is $H_k(0)=\delta_{k,0}$ and the
boundary conditions are $H_0=1$ and $H_{-1}=0$.  The first term on the
right-hand side accounts for gain due to node addition and the next
two terms describe changes due to node deletion.  The factor $k$
reflects that deletion of a node of height $k$ reduces the heights of
all of its descendants by one. Again, the addition and the deletion
rates are normalized by the total number of nodes and the total number
of links, respectively. One can verify that the total number of nodes
$N=\sum_k H_k$ satisfies \hbox{$dN/dt=r-1$}.

Following the above analysis of the in-component size distribution,
we study the long time asymptotic behavior of $h_k$, the fraction of
nodes with height $k$, using the transformation
\begin{equation}
\label{hk-def}
H_k=Nh_k.
\end{equation}
Normalization implies $\sum_k h_k=1$ and the boundary
condition is now $h_0=0$ since $H_0=1$. Let us substitute
(\ref{hk-def}) into the rate equation (\ref{Hkt-eq}). The distribution
function $h_k$ satisfies the recursion relation
\begin{equation}
\label{hk-eq}
r(h_k-h_{k-1})+(h_{k+1}-h_k)=(k+1)h_{k+1}-kh_k.
\end{equation}
This equation was obtained by adding $H_{k+1}$ to both side of
(\ref{Hkt-eq}).

In this case, it is convenient to use the exponential generating
function, defined via
\begin{equation}
\label{Gz-def}
G(z)=\sum_{k=0}^\infty h_k e^{kz}.
\end{equation}
Multiplication of the recursion relation (\ref{hk-eq}) by $e^{kz}$ and
summation over $k$ leads to the first order ordinary differential
equation
\begin{equation}
\frac{dG}{dz}=G(z)(1+re^z).
\end{equation}
The generating function is found by integration of this equation 
subject to the boundary condition $G(0)=1$,  
\begin{equation}
\label{Gz}
G(z)=e^{z+r(e^z-1)}.
\end{equation}
Using the definition \eqref{Gz-def} we find that the height
distribution function is Poissonian
\begin{equation}
\label{hk-sol}
h_k=e^{-r}\frac{r^{k-1}}{(k-1)!}.
\end{equation}

The moments of the height distribution \hbox{$m_n=\sum_k k^nh_k$} can
be found by differentiation of the generating function (\ref{Gz}).
For instance, $m_1=G'(0)=1+r$, and therefore the average node height
is equal to the average in-component size, $\langle k\rangle=1+r$.
Similarly, fluctuations in the height are obtained from the second
moment $m_2=c''(0)=r^2+3r+1$, and hence, the variance is
\begin{eqnarray}
\langle k^2\rangle-\langle k\rangle^2=r.
\end{eqnarray}
This variance is sufficient to characterize the behavior in the
vicinity of the average, where the Poissonian height distribution
becomes Gaussian.

We can immediately deduce a number of characteristics using the height
distribution.  There are $H_k$ nodes at height $k$ with incoming links
from the $H_{k+1}$ nodes at height $k+1$. Thus, the average in-degree
$I_k$ of nodes at height $k$ is $I_k=H_{k+1}/H_k$. Using $H_0=1$ and
(\ref{hk-sol}) we find
\begin{equation}
\label{Ik}
I_k=
\begin{cases}
Ne^{-r}&k=0;\\
r\,k^{-1}&k>0.
\end{cases}
\end{equation}
The second result shows that the average degree is inversely
proportional to the height.

The root, with its macroscopic degree, is the most connected node
whereas all other nodes have a microscopic degree. Hence, the root is 
a giant hub.  The degree of the root is maximal for marginally growing
networks, $\lim_{r\to 1}I_0=Ne^{-1}$, and it decreases indefinitely as
$r$ increases.  Therefore, deletion must be responsible for this
condensation phenomena. Indeed, since the attachment rate to each node
is $1/N$, addition can be responsible for at most $\ln N$ of the
$Ne^{-r}$ connections as is the case for random recursive trees
\cite{kr1,mh,drmota}. We conclude that emergence of the giant hub is
due to deletion.

Consider now the highest node.  The maximal height $k_{\rm max}$ is
determined from the extreme statistics criterion $H_{k_{\rm max}}\sim
1$. Using the Poisson distribution (\ref{hk-sol}) and the Stirling
formula $k!\sim (k/e)^k$, the highest node grows very slowly with the
network size, $k_{\rm max}\approx {\ln N}/{\ln(\ln N)}$.  Remarkably,
the maximal height is independent of the addition rate $r$.

Asymptotically, the diameter of the network $D$ is twice the maximal
height
\begin{equation}
\label{diameter}
D \approx 2\,\frac{\ln N}{\ln(\ln N)}.
\end{equation}
For strictly growing random trees, the diameter exhibits a logarithmic
growth independent of the attachment mechanism, $D\sim \ln N$
\cite{kr1}.  However, the diameter can be affected by the topology.
If every new node links to $m$ existing nodes, this strictly growing
network is no longer a tree (for $m\geq 2$) and $D\sim\ln N/\ln \ln
N$. This result \cite{br} is again very robust, namely it holds for
strictly growing networks with reasonable attachment mechanisms and
for all $m\geq 2$. Equation \eqref{diameter} shows that deletion
qualitatively affects the growth of the diameter.  Surprisingly,
despite their distinct topologies these strictly growing networks and
the addition-deletion networks have similar diameters.

Finally, we address the average distance between two nodes, $\langle
l\rangle$.  Since the degree of the root is proportional to the total
number of nodes $N$, a path connecting two randomly selected nodes
almost surely includes the root. As a result, the average distance is
asymptotically equal to twice the average height $\langle
l\rangle=2(1+r)$.

\section{The Degree Distribution}

In contrast with the in-component size distribution and the height
distribution, it is impossible to construct closed equations for the
degree distribution and we perform an approximate analysis.

When a node with in-degree $j$ is deleted, the in-degree of the parent
$i$ is augmented by $j-1$. This is simply the aggregation process
\cite{fl}
\begin{equation}
\label{agg}
(i,j)\buildrel i/(NL)\over \longrightarrow i+j-1.
\end{equation}
The aggregation rate is proportional to the in-degree of the parent
node because any of the daughter nodes can be deleted.  The
normalization factor $NL$ reflects that the probability of picking the
deleted node is inversely proportional to the number of nodes and the
probability of picking the parent node is inversely proportional to
the number of links.

Let $A_k(t)$ be the average number of nodes with in-degree $k$ at time
$t$ (the out-degree is always one).  The total number of nodes and the
total number of links are given by $N=\sum_k A_k$ and $L=\sum_k kA_k$,
respectively.  Assuming that the degrees of neighboring nodes are
completely uncorrelated, the quantity $A_k$ obeys the nonlinear 
rate equation \cite{mvs,bk1}
\begin{eqnarray}
\label{Ak-eq}
\frac{dA_k}{dt}&=&r\delta_{k,0}+\frac{r}{N}(A_{k-1}-A_k)\\
&+&\frac{1}{NL}\left[\sum_{i+j=k+1}iA_iA_j-(kN+L)A_k\right].\nonumber
\end{eqnarray}
The initial condition is $A_k(0)=\delta_{k,0}$ and the boundary
condition is $A_{-1}(t)=0$. The first three terms represent addition
events. These terms are linear and thus, exact. The last three terms
account for deletion events. These terms are non-linear as they
represent two node interactions. Moreover, the gain term has a
convolution structure, reflecting the aggregation process \eqref{agg}.
Of course, the aggregation process becomes exact in the limit
$r\to\infty$ where deletion becomes irrelevant.  By summation of the
rate equation \eqref{Ak-eq} we recover $dN/dt=dL/dt=r-1$.

Following our above analysis, we study the fraction of nodes with
in-degree $k$, $a_k$. Making the transformation $A_k=Na_k$, the degree
distribution satisfies
\begin{equation}
\label{ak-eq}
(k+2r)a_k=r\delta_{k,0}+r a_{k-1}+\sum_{i+j=k+1}ia_ia_j
\end{equation}
We analyze this hierarchy of equations using the generating function $
U(x)=\sum_k a_k x^k$.  The normalizations $\sum_k a_k=\sum_k ka_k=1$
imply \hbox{$U(1)=U'(1)=1$}. By multiplying the recursive equations
\eqref{ak-eq} by $x^k$ and summing over $k$, we find that the
generating function obeys the nonlinear ordinary differential equation
\begin{equation}
\label{Uz-eq}
(x-U)\frac{dU}{dx}=r(x-2)U+r.
\end{equation}
This equation is consistent with the normalization conditions
\hbox{$U(1)=U'(1)=1$}. 

\begin{figure}[t]
\includegraphics[width=0.45\textwidth]{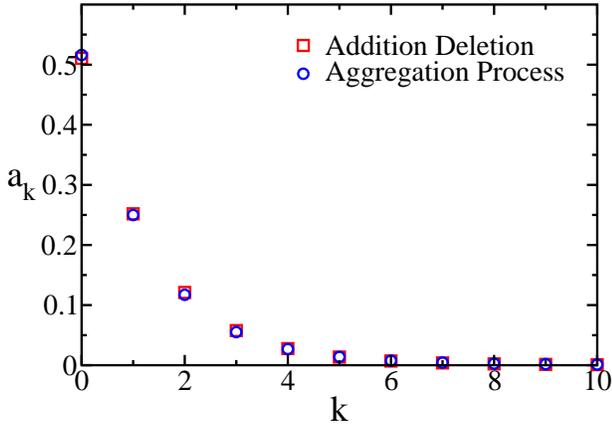}
\caption{The degree distribution for the aggregation process and for
the addition-deletion network. Shown are results of one Monte Carlo
simulation run with $10^7$ nodes and $r=4$. Roughly 90\% of the nodes
have a degree smaller or than \hbox{$k=10$}.}
\label{fig-ak}
\end{figure}

The moments of the degree distribution defined by $m_n=\sum_k a_k k^n$ are
obtained by successive differentiation of the generating function. In
particular, the high-order moments grow rapidly although they do
remain finite (appendix B)
\begin{equation}
\label{mn}
m_n\sim \left[\frac{r}{(r-1)^2}\right]^n (n!)^2,
\end{equation}
as $n\to\infty$. Using these asymptotics we deduce (see Appendix B)
that the tail of the degree-distribution decays as a stretched
exponential
\begin{equation}
\label{ak-tail}
a_k\sim e^{-\lambda \sqrt{k}}
\end{equation}
with $\lambda=[2(r-1)]/\sqrt{r}$. 

How well does this approximation compare with addition-deletion
networks? In addition-deletion networks, the root has a macroscopic
degree. In other words, it is a condensate that contains a finite
fraction of all nodes. In the aggregation process (\ref{agg}) all
clusters are finite, as seen from $U'(1)=\sum_k ka_k=1$
\cite{bk1}. This is a consequence of the fact that all nodes are
treated in the same way. Introducing a fixed node that can not be
deleted should significantly improve the approximation.

We also examined the degree distribution using Monte Carlo simulations
of the aggregation process (\ref{agg}) and the addition-deletion
process.  We find that when the degree is very small, the aggregation
process approximates the addition-deletion process very well (Figure
\ref{fig-ak}). For small degree this agreement holds independent of
$r$. However, this approximation is poor when the degree is large.

\begin{figure}[t]
\includegraphics[width=0.45\textwidth]{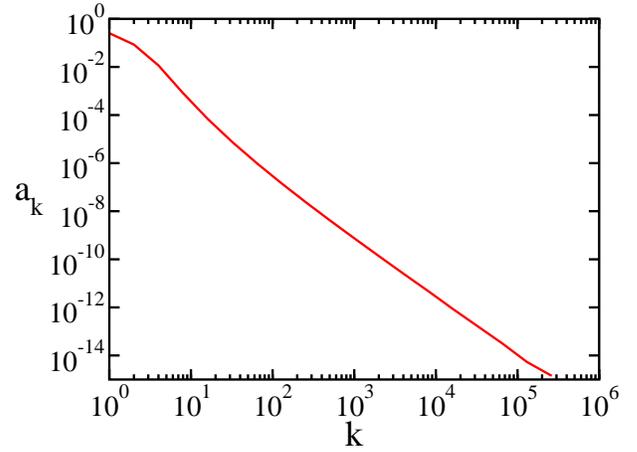}
\caption{The tail of the degree distribution for the addition-deletion
 network. The simulation results were obtained from $10^3$ independent
 realizations in a network with $10^7$ nodes and $r=4$. The decay
 exponent is $\gamma=2.45\pm 0.05$}
\label{fig-tail}
\end{figure}

For addition-deletion networks, the degree distribution
follows a power law behavior (Figure \ref{fig-tail})
\begin{equation}
a_k\sim k^{-\gamma}
\end{equation}
as $k\to\infty$. Hence, degree-degree correlations are negligible for
small degrees but substantial for large degrees. Indeed, highly
connected nodes are long lived and thus, involve strong memory.

Addition-deletion network can be simulated using an efficient
algorithm that is linear in the network size. Throughout the growth
process we keep track of all nodes, whether they have been deleted
(passive nodes) or not (active nodes). In an addition event, an active
node is added and it links to a randomly selected active node. In a
deletion event, an active node is marked passive. When the required
network size is reached, passive nodes are removed in reverse
chronological order to preserve the ancestral relations.

Simulations of the aggregation process are performed as follows. The
system consists of aggregates, each with a given size. In an addition
step, an aggregate of size zero is added and simultaneously, the size
of another, randomly chosen aggregate increases by one. In a deletion
step one aggregate is chosen with probability proportional to its size
and it merges with another randomly chosen aggregate according to
(\ref{agg}).

\section{Summary}

We have studied networks which undergo a biased growth --- new nodes
are added at a certain rate and old nodes are deleted at a smaller
rate in such a way that the ancestry is preserved.  This growing
network model is relevant to biological and technological networks
that preserve ancestral relations between nodes.  Our main conclusion
is that it is possible to characterize the structure of such networks
analytically. For instance, the in-component size distribution has a
power-law tail and the characteristic exponent varies continuously
with the addition rate.  We have also shown that the height
distribution obeys Poisson statistics. Additional properties including
the fraction of dangling nodes and the average degree as a function of
height have been obtained as well.

Interestingly, the addition-deletion process results in a giant hub
that is connected to a finite fraction of the nodes in the
system. This fraction decreases to zero as deletion becomes weaker,
showing that the deletion process causes this condensation phenomena.
 
We have also seen that the degree distribution evolves by an
aggregation process since the parent node inherits all incoming links
of a deleted node. Treating the underlying aggregation process as
completely random is valid only at small degrees because degree-degree
correlations become significant at large degrees.  Constructing closed
equations for this aggregation process from which the degree
distribution can be calculated is an outstanding challenge.

\noindent{\bf Acknowledgments.}  We acknowledge financial support from
DOE grant DE-AC52-06NA25396 and NSF grant CHE-0532969.

\appendix

\section{Derivation Eq.~(\ref{alpha})}
\label{Alpha}

The tail of the in-component size distribution can be obtained from
the $x\to 1$ asymptotics of the generating function $F(x)$.  Let us
consider for example the power-law tail $c_k\simeq B k^{-2}$ as
$k\to\infty$.  In this case, the leading behavior of the derivative of
the generating function is
\begin{eqnarray*}
\frac{dF(x)}{dx}=\sum_k kc_kx^{k-1}\simeq B \sum_k k^{-1}x^k=
-B \ln(1-x)
\end{eqnarray*}
in the limit $x\to 1$. Generally, the power-law decay (\ref{beta})
implies
\begin{equation}
\label{deriv-a}
\frac{d^{\beta-1}F(x)}{dx^{\beta-1}}=-B \ln(1-x)
\end{equation}
as $x\to 1$, for any positive integer $\beta$.

To evaluate this derivative, we rewrite the generating function
(\ref{Fx-sol-a}) using the transformation $z=(\beta-1)(1-y)$ 
\begin{equation}
\label{Fx-sol-c}
F(x)\!=\!r^\beta\,\frac{x(1-x)^{\beta-1}}{(rx-1)^\beta}
\int_{(\beta-1)(1-x)}^1\!\!\! dz\,\frac{(1-z)^{\beta-1}}{z^\beta}.
\end{equation}
For arbitrary positive integer $\beta$, the integral is evaluated by
performing $\beta$ successive integration by parts
\begin{eqnarray}
\label{Fx-sol-d}
F(x)\!=\!\frac{x P_{\beta-1}(x)}{(rx-1)^\beta}
\!+\!(-r)^\beta\frac{x(1-x)^{\beta-1}}{(rx-1)^\beta}\,\ln(1-x)
\end{eqnarray}
where $P_n(x)$ is a polynomial of degree $n$. 

Differentiating the generating function (\ref{Fx-sol-d}) $\beta-1$
times and noting that the second term dominates in the $x\to 1$ limit,
we obtain
\begin{equation}
\label{deriv-b}
\frac{d^{\beta-1}F(x)}{dx^{\beta-1}}
\simeq -(\beta-1)!\left(\frac{r}{r-1}\right)^\beta\, \ln(1-x).
\end{equation}
By comparing this expression with (\ref{deriv-a}) we obtain the
coefficient \hbox{$B=\Gamma(\beta)(\beta-1)^\beta$} for all integer
$\beta\geq 2$. Finally, we use analytic continuation and extend this
result to arbitrary $\beta$.

\section{Derivation of (\ref{mn}) and (\ref{ak-tail})} 
\label{Aggr}

We obtain the large-$k$ behavior of the degree distribution using the
$x\to 1$ behavior of the generating function $U(x)$.  Differentiating
equation (\ref{Uz-eq}) $n$ times and setting $x=1$ gives a recursion
relation for the derivatives $U_n=d^nU(x)/dx^n|_{x=1}$
\begin{eqnarray*}
U_n=\frac{r\,n(n+r)}{(r-1)^2}U_{n-1}+\frac{1}{r-1}\sum_{m=2}^{n-3}{n\choose
m}U_{n-m}U_{m+1}
\end{eqnarray*}
The first two coefficients are $U_0=U_1=1$.  Only the first term in
the recursion equation is relevant asymptotically and therefore,
\begin{equation}
\frac{U_n}{U_{n-1}}\to \frac{r}{(r-1)^2}\,n^2
\end{equation}
as $n\to\infty$. Eq.~(\ref{mn}) is obtained by noting that the
coefficients
\begin{equation*}
U_n=\sum_{k\geq 0} k(k-1)\ldots(k-n+1)a_k
\end{equation*}
are asymptotically equivalent to the moments, $U_n\sim m_n$. 

The large-$n$ behavior of $m_n$ is computed as follows:
\begin{eqnarray}
m_n&\sim& \int dk\, k^n a_k \nonumber\\
&\sim& \int dk\,\exp\left(n\ln k-\lambda\sqrt{k}\right)\nonumber\\
   &\sim& \exp\Big[2n\ln(2n/\lambda e)\Big]\nonumber\\
\label{mn1}
   &\sim& \left(\frac{2}{\lambda}\right)^{2n}(n!)^2.
\end{eqnarray}
In the first line we replaced the summation by integration, and in the
second we used the presumed asymptotic $a_k\sim e^{-\lambda\sqrt{k}}$.
The third line was obtained by using the steepest descent method ---
the function \hbox{$f(k)=n\ln k-\lambda\sqrt{k}$} is maximal at
\hbox{$k_*=(2n/\lambda)^2$}. The fourth line was obtained using the
Stirling formula \hbox{$n!\sim (n/e)^n$}.  Finally, by comparing
equations (\ref{mn1}) and \eqref{mn}, we obtain the tail behavior
(\ref{ak-tail}) \cite{general}.

\end{document}